\documentclass[preprint,authoryear,12pt,3p,a4paper]{elsarticle1}
\usepackage[utf8x]{inputenc}
\usepackage{units}
\usepackage{amsmath,amssymb,amstext}
\usepackage{graphicx}

\begin{document}

\title{Empirical Evidence for the Structural Recovery Model}

\author{Alexander Becker}
\address{Faculty of Physics, University of Duisburg-Essen, \\ Lotharstrasse 1, 47048 Duisburg, Germany; \\
email: alex.becker@uni-duisburg-essen.de}

\author{Alexander F.~R.~Koivusalo}
\address{Koivusalo Capital, Portfolio Analytics, \\ \"Ostra F\"orstadsgatan 21, 21131 Malm\"o, Sweden; \\ 
email: alexander.koivusalo@gmail.com} 

\author{Rudi Sch\"afer}
\address{Faculty of Physics, University of Duisburg-Essen, \\ Lotharstrasse 1, 47048 Duisburg, Germany; \\[1ex] 
email: rudi.schaefer@uni-duisburg-essen.de}

\begin{abstract}

While defaults are rare events, losses can be substantial even for credit portfolios with a large number of contracts. Therefore, not only a good evaluation of the probability of default is crucial, but also the severity of losses needs to be estimated. The recovery rate is often modeled independently with regard to the default probability, whereas the Merton model yields a functional dependence of both variables. We use Moody's  Default and Recovery Database in order to investigate the relationship of default probability and recovery rate for senior secured bonds.
The assumptions in the Merton model do not seem justified by the empirical situation. Yet the empirical dependence of default probability and recovery rate is well described by the functional
dependence found in the Merton model.
\vspace*{2ex}
\end{abstract}


\maketitle


\section{Introduction}
\noindent
The recent debt crises demonstrate that a detailed understanding of credit risk is of crucial importance. While defaults are generally rare events, losses can be substantial even for large credit portfolios. 

In the past, much effort has been devoted to assess the probability of default with elaborate models. While the original Merton model, see \cite{Merton1974}, performs rather poorly at predicting the default risk, improved structural models and reduced-form models are more successful in this regard, demonstrated in \cite{DuffieSingleton1999}, \cite{Shumway2001}, \cite{Chava2004}.
As these models focus on the probability of default, they tend to lose sight of the severity of default events, that is, the loss given default and the corresponding recovery rate. \cite{Altman2004} point out that, instead, the recovery rate is usually modeled independently. For an example, see \cite{AsvanuntStaal2009a} and \cite{AsvanuntStaal2009b}. 

Recent works show a negative correlation of default probability and recovery rate, as in \cite{Frye2000a}, \cite{Frye2000b}, \cite{Altman2005}, and \cite{Acharya2007}. The Basel II regulatory capital framework also requires this aspect to be included, compare \cite{BaselII}. 
While the negative correlation can be successfully introduced into reduced-form models, as \cite{Chava2008} and \cite{BadeRoeschScheule2011} demonstrated, this lacks a deeper motivation. In the original Merton framework, however, the inverse relationship is naturally embedded. \cite{KoivusaloSchaefer2011a} derived the functional dependence of default and recovery rates in the structural model. 

Detailed empirical studies regarding the recovery rate have not been feasible due to scarce historical data, though. This has changed with the financial turmoil of the last decade. A large number of default events and recovery rates were recorded. This renders an extensive investigation of the relationship of default probability and recovery rate possible.

In our empirical study, we find that the structural recovery describes the credit data well. This allows to extend existing default models in such a way that the implicit recovery information is taken into account.

The paper is organized as follows. In section 2, we will present a short review of the Merton model. In section 3, we will briefly explain Moody's Default and Recovery Database. 
We will discuss how we built credit portfolios, and how we computed their default and recovery rates. The empirical results are given in section 4. We will summarize our findings with a conclusion in section 5.

\section{Merton Model}
\noindent
In the framework of the Merton model, the liabilities of a company are modeled as a single zero-coupon debt. We want to examine a portfolio consisting of credit contracts with $K$ companies which we build at $t=0$. Each company $k$ has a debt with face value $F_k$ due at time $T$, the maturity time. In this model, a company will default if the market value of its assets, $V_k(T)$, is less than $F_k$ when the bond matures.

In order to assess the credit risk, we need to investigate the market value $V_k(t)$. The value of the assets of the company are amenable to stochastic modeling. Let $\widetilde{p_k}(V_k)$ be the probability density function of the market value of the company at maturity. The probability of default ($\mathrm{PD}$) is then given by 
  \begin{equation}
    \mathrm{PD}_k = \int_0^{F_k} \widetilde{p_k}(V_k) \,\mathrm{d}V_k\,.
    \label{eq:PD}
  \end{equation}
The recovery rate ($\mathrm{RR}$) is simply $V_k(T)/F_k$ for $V_k(T)<F_k$. Hence the expected recovery rate for the same stochastic process reads
  \begin{equation}
    \langle \mathrm{RR}_k\, \rangle = \frac{1}{\mathrm{PD}_k}\,\int_0^{F_k} \frac{V_k}{F_k}\,\widetilde{p_k}(V_k) \,\mathrm{d}V_k\,.
    \label{eq:RR}
  \end{equation}
\cite{KoivusaloSchaefer2011a} derived the results for Eqs.~\eqref{eq:PD} and \eqref{eq:RR} for a correlated diffusion process with drift $\mu$, volatility $\sigma$
and correlation $c$,
  \begin{equation}
   \frac{\mathrm{d}V_k}{V_k} = \mu \mathrm{d}t + \sqrt{c}\,\sigma\,\mathrm{d}W_m + \sqrt{1-c}\,\sigma\,\mathrm{d}W_k\,.
  \end{equation}
The Wiener processes $\mathrm{d} W_k$ and $\mathrm{d} W_{\rm m}$ describe the idiosyncratic and the market fluctuations, respectively.
\cite{KoivusaloSchaefer2011a} find for this underlying process
  \begin{equation}
    \langle \mathrm{RR}\,(\mathrm{PD}) \rangle =
       \frac{1}{\mathrm{PD}} \exp\left(-B\,\Phi^{-1}(\mathrm{PD})+\frac{1}{2}B^2\right)\;\Phi\left(\Phi^{-1}(\mathrm{PD})-B\right),
    \label{eq:PDRR}
  \end{equation}
with the compound parameter 
  \begin{equation}
    B = \sqrt{(1-c)\sigma^2\,T}\,,
    \label{eq:B}
  \end{equation}
where $\Phi$ is the cumulative standard normal distribution. $\sigma$ and $c$ are the aforementioned volatility and correlation, respectively. 
It is assumed that $\sigma$ and $c$ have the same constant value for each company, regardless of $T$.
We also consider $V_{0,k}$, the initial value of the assets, and $F_k$ to be the same for each company, namely $V_0$ and $F$.

\begin{figure}[ht]
    \centering
    \includegraphics[scale=1]{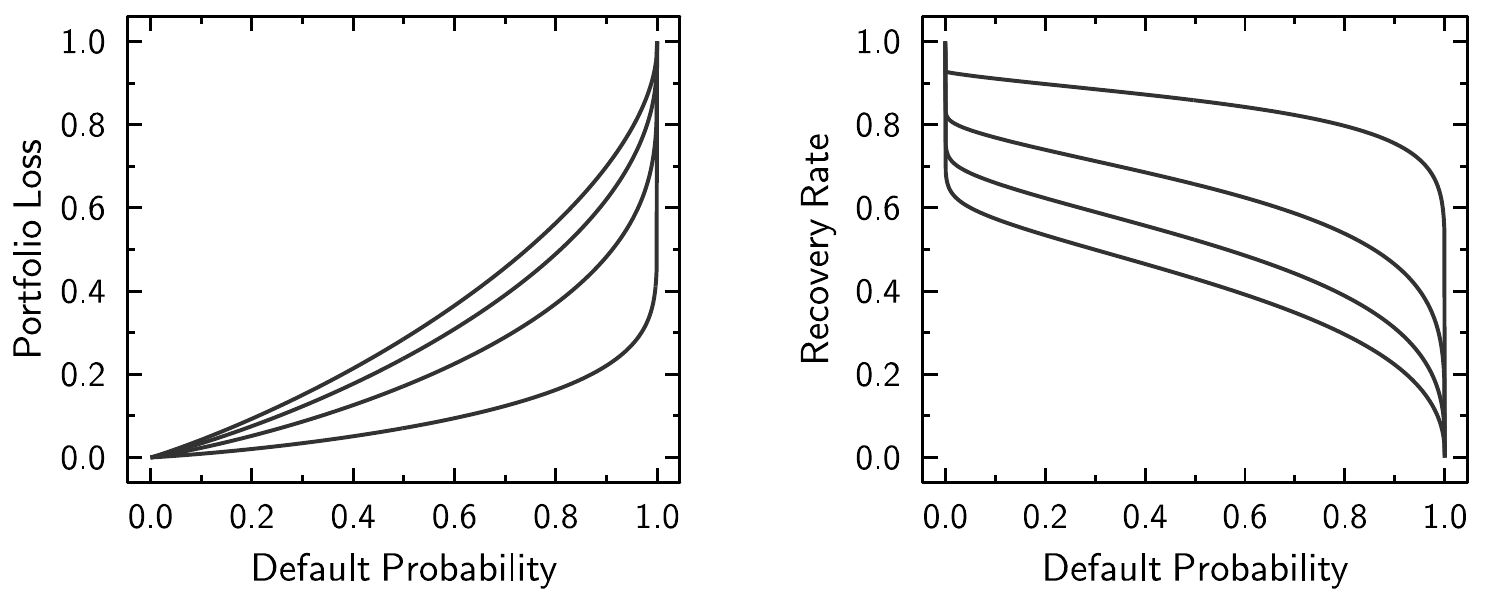}
    \caption{Dependence of $\langle L(\mathrm{PD}) \rangle$ and $\langle \mathrm{RR}\,(\mathrm{PD}) \rangle$ on parameter $B$, respectively.
    In the left plot the parameter takes the values $B = 0.2, 0.6, 1.0, 1.4$, from bottom to top; vice versa in the right plot.}
    \label{fig:PDPDRR}
  \end{figure}

The recovery rate and the expected portfolio loss $\langle L \rangle$ are connected through
  \begin{equation}
    \langle L(\mathrm{PD}) \rangle =  \mathrm{PD} \cdot (1 - \langle \mathrm{RR}\,(\mathrm{PD}) \rangle)\,.
  \end{equation}
Inserting Eq.~\eqref{eq:PDRR} for the recovery rate yields
  \begin{equation}
    \langle L(\mathrm{PD}) \rangle =
       \mathrm{PD} - \exp\left(-B\,\Phi^{-1}(\mathrm{PD})+\frac{1}{2}B^2\right)\;\Phi\left(\Phi^{-1}(\mathrm{PD})-B\right)\,.
    \label{eq:PDLGD}
  \end{equation}
Figure \ref{fig:PDPDRR} illustrates the relationships arising from Eqs.~\eqref{eq:PDRR} and \eqref{eq:PDLGD}.
We find that the expected portfolio loss increases with the default probability, while the recovery rate decreases. 
This inverse relation between default and recovery rates is in accordance with earlier findings, see, for example, \cite{Frye2000a}, \cite{Frye2000b}, \cite{Altman2005}, \cite{Acharya2007}, and \cite{BadeRoeschScheule2011}. 
However, the dependence between these two quantities is not linear as the use of the correlation coefficient might suggest.
In the following we will examine this aspect in detail. In particular, we will compare the empirical relation of these quantities with the structural recovery rate given by Eq.~\eqref{eq:PDRR}.

\section{Moody's Default and Recovery Database}
\noindent
For our investigation, we use Moody's Default and Recovery Database\footnote{For details see \texttt{www.moodysanalytics.com}} (DRD). 
According to \cite{MoodyTechSpec}, the DRD contains data for 40,000 global corporate and sovereign entities, including more than 417,000 debts. The database is particularly useful because it includes recovery rates for most default events.

Moody's have assigned a credit rating to each current bond issuer listed in the database. This rating is based on the estimated default probability of the issuer in a given period of time.
This allows us to sort the bond issuers by their estimated default risk and, thus, to build credit portfolios of similar types. In our analysis, we consider the default probability of these portfolios for different maturity times and the corresponding recovery rates.

\subsection{Default Probabilities}
\noindent
In order to calculate the default probability, we have to first set up a credit portfolio in a reasonable matter. In the following, we will refer to our credit portfolio as a cohort.
The cohort is characterized by the following parameters. We have to select a date for building the cohort. In the Merton model, this date corresponds to $t=0$, the beginning of the term of the bonds. In reality, the bonds or loans have usually been issued beforehand.
We then select the members of the cohort by their rating at $t=0$. Additionally, we choose the seniority of the debt. At date $t=T$, we investigate the condition of the debt issues. 
When we compare the empirical results to the Merton model, this date will correspond to the maturity date in the model. However, it does not necessarily coincide with the actual maturities of the debts in the DRD. Therefore, we call $T$ the model-maturity. Furthermore, we have to keep in mind that the debt structure of companies is often more complicated than in the Merton model.

There are three possible outcomes for a member of the cohort. First, the issuer might default prior to $T$, the model-maturity.
Second, the issuer might  be withdrawn; i.e. the debt has gone private, or it has been fully paid back by time $T$. Third, neither might happen, to which we will refer as survival.
In order to calculate the actual default rate of the portfolio, we have to take the withdrawals into account. The empirical data indicates that the frequency of defaults and withdrawals is not correlated.

Therefore, the default rate of the cohort can be calculated as follows. The cohort has $n_{\mathrm{C}}$ members, each is subject to default or withdrawal with probabilities $\mathrm{PD}$ and $p_{\mathrm{W}}$, respectively. We can only observe defaults of issuers that have not been withdrawn. Hence, the number of defaults $n_{\mathrm{D\overline{W}}}$ is 
  \begin{equation}
    n_{\mathrm{D\overline{W}}} = n_{\mathrm{C}} \, \big(1-p_{\mathrm{W}}\big) \, \mathrm{PD} \;=\; n_{\mathrm{C}} \, \Big(1-\frac{n_{\mathrm{W}}}{n_{\mathrm{C}}}\Big) \, \mathrm{PD}\,.
  \end{equation}
From this, we get
  \begin{equation}
    \mathrm{PD} = \frac{n_{\mathrm{D\overline{W}}}}{n_{\mathrm{C}}-n_{\mathrm{W}}}\,.
    \label{eq:PDDB}
  \end{equation}
Note that we are solely interested in the outcome at model-maturity. Therefore, we treat any event between $t=0$ and $t=T$ as though it happened at time $T$.

\subsection{Recovery Rates}
\noindent
After a default has occurred the debt is still traded on the market. Therefore it is possible to assess the recovery rate by the trading price 30 days after default;
this is the market value of the debt as a percentage of the face value. We find this information in the DRD. In addition, the ultimate recovery rates are included in the database if available.

In this paper, we focus on the recovery rates based on the trading prices. We do so for two reasons. First, issuers listed in the ultimate recovery database have commonly undergone the Chapter 11 process. The economic conditions for the business of a company are subject to change during this period of time. Second, and even more important, the sample is much larger if we the use the trading prices. For example, Moody's does neither include financial institutions nor issuers with an exposure at default of less than \$50 million in its ultimate recovery database, as explained in \cite{MoodyUltiRec}.

\section{Empirical Results}
\subsection{Correlation between Default Probability and Recovery Rate}\label{sec:Corr}

\noindent
We measure the default probability and the recovery rate for debts with a model-maturity of one year from January 1st, 2000, to December 31st, 2010.
The first term starts January 1st, 2000, and ends December 31st, 2000. The second term starts February 1st, 2000, and ends January 31st, 2001.
Accordingly, the last term starts January 1st, 2010, and ends December 31st, 2010.

  \begin{figure}[!ht]
    \centering
    \includegraphics[scale=1]{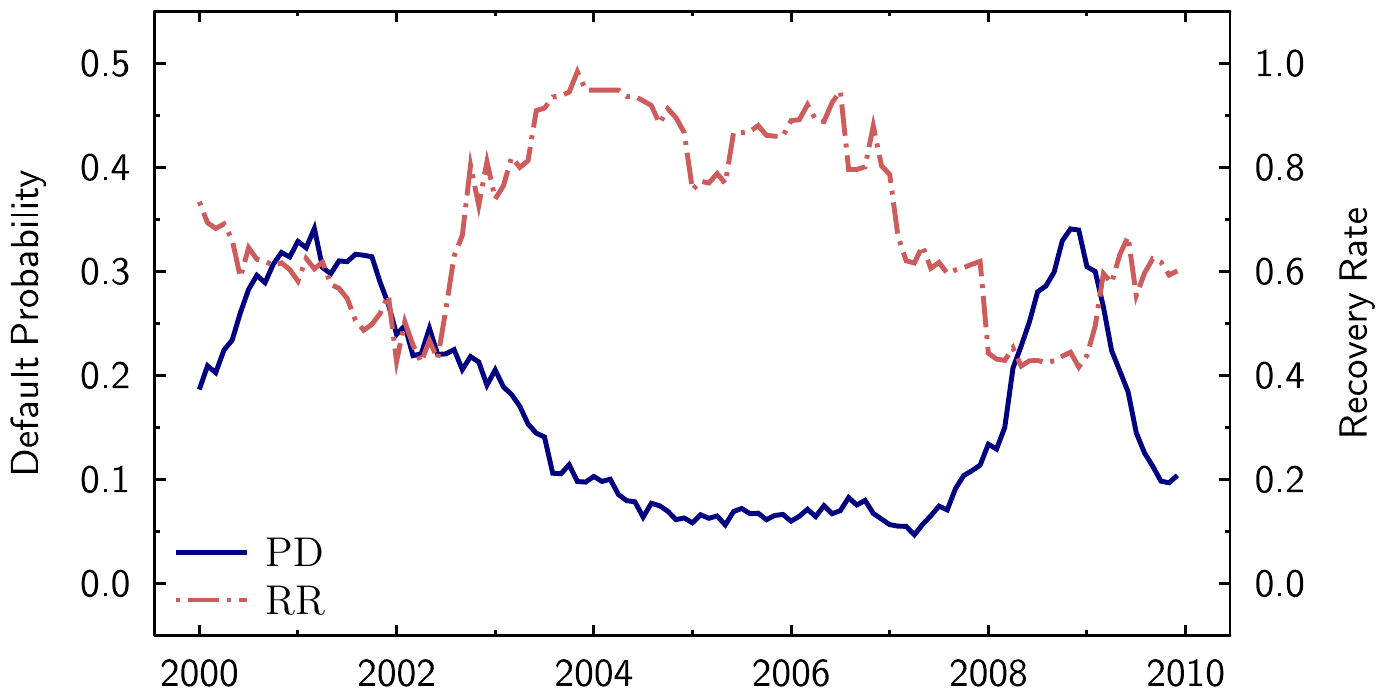}
    \caption{Default probability and recovery rate for senior secured bonds with ratings $\mathsf{Caa1}$, $\mathsf{Caa2}$ and $\mathsf{Caa3}$. Both quantities were evaluated on one year intervals. The time axis indicates the starting dates of these intervals. The correlation of default probabilities and recovery rates is $-0.679$.}
    \label{fig:CorrCaa}
  \end{figure}

We evaluate the default probability and recovery rate for senior secured bonds with ratings $\mathsf{Caa1}$, $\mathsf{Caa2}$ and $\mathsf{Caa3}$. We present the results in Fig.~\ref{fig:CorrCaa}.
They are strongly negatively correlated with correlation coefficient $-0.679$. Data sets with other ratings share the illustrated negative correlation.
It is worth pointing out that the default probability and recovery rate are inversely related for senior unsecured bonds as well. In this case, we find a correlation coefficient of $-0.639$.  We find the same behavior for longer model-maturities.

\subsection{Dependence of Recovery Rate on Default Probability}
\noindent
We recall that we use the data in the context of the Merton model. In a nutshell, we select bonds with a certain rating for our portfolio, and we assume that they are issued on the same date the portfolio is set up. Furthermore, we determine the outcome of the bond at its maturity time in the Merton framework by the state of the bond that is registered in the DRD at that particular date, that is, at model-maturity $t=T$.

Thus, we face two restrictions: the portfolio size (the number of companies carrying a given rating at a given date) and the number of default events with complete information (the number of defaults with recovery information that have occurred during the model-maturity of the bonds). The prevalent problem is the latter. For one-year model-maturities, e.g., we find that on average less than one default with recovery information is recorded for ratings $\mathsf{Ba3}$ and better.

In our evaluation, we hence investigate portfolios of lower initial ratings, namely $\mathsf{B1}$, $\mathsf{B2}$, $\mathsf{B3}$, $\mathsf{Caa1}$, $\mathsf{Caa2}$ and $\mathsf{Caa3}$.
These are the ratings which are generally referred to as risky and highly speculative. In order to set up the cohorts, we then proceed as described in section \ref{sec:Corr}.

\begin{figure}[!ht]
    \centering
    \includegraphics[scale=1]{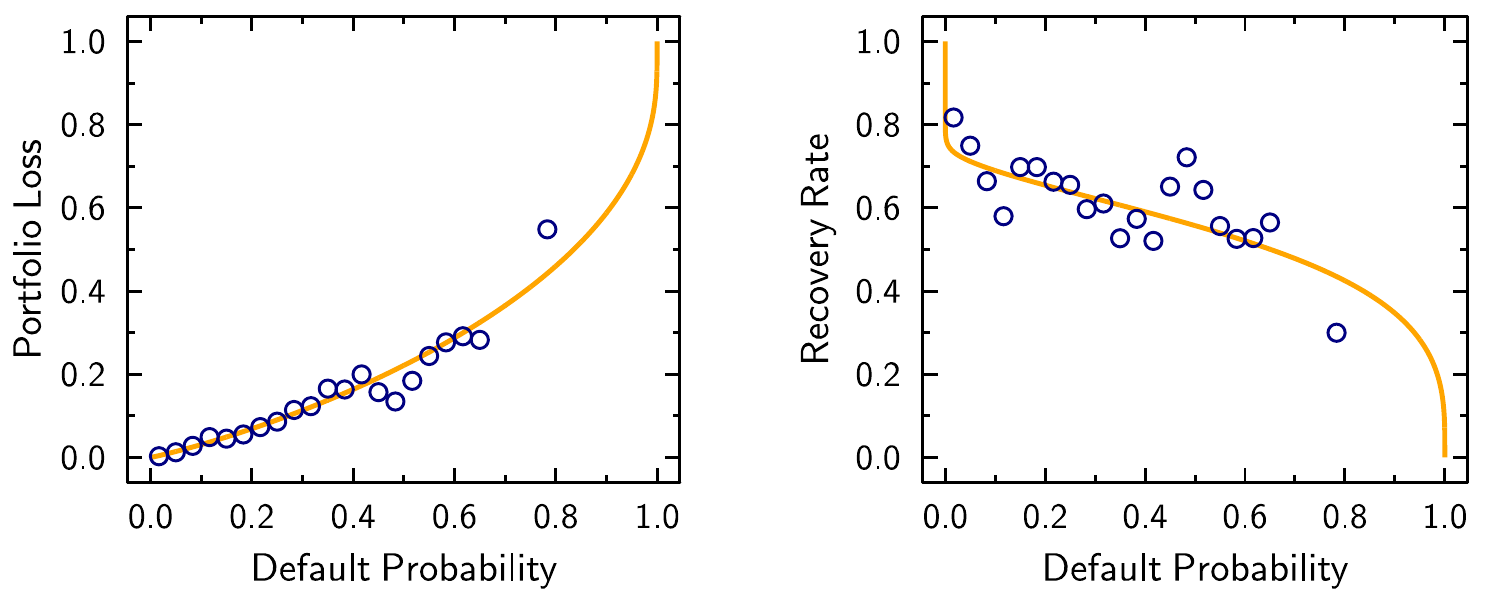}
    \caption{Empirical results from 2000 to 2010 for two-year model-maturities.
    The left plot shows the portfolio loss $\langle \mathrm{L}(\mathrm{PD})
\rangle$, and the right plot shows the recovery rate $\langle
\mathrm{RR}(\mathrm{PD}) \rangle$. The orange line shows the respective
analytical curves with $B = 0.882$. The parameter has been fitted to the
portfolio loss data.}
    \label{fig:RRLGD2}
  \end{figure}

\begin{figure}[!ht]
    \centering
    \includegraphics[scale=1]{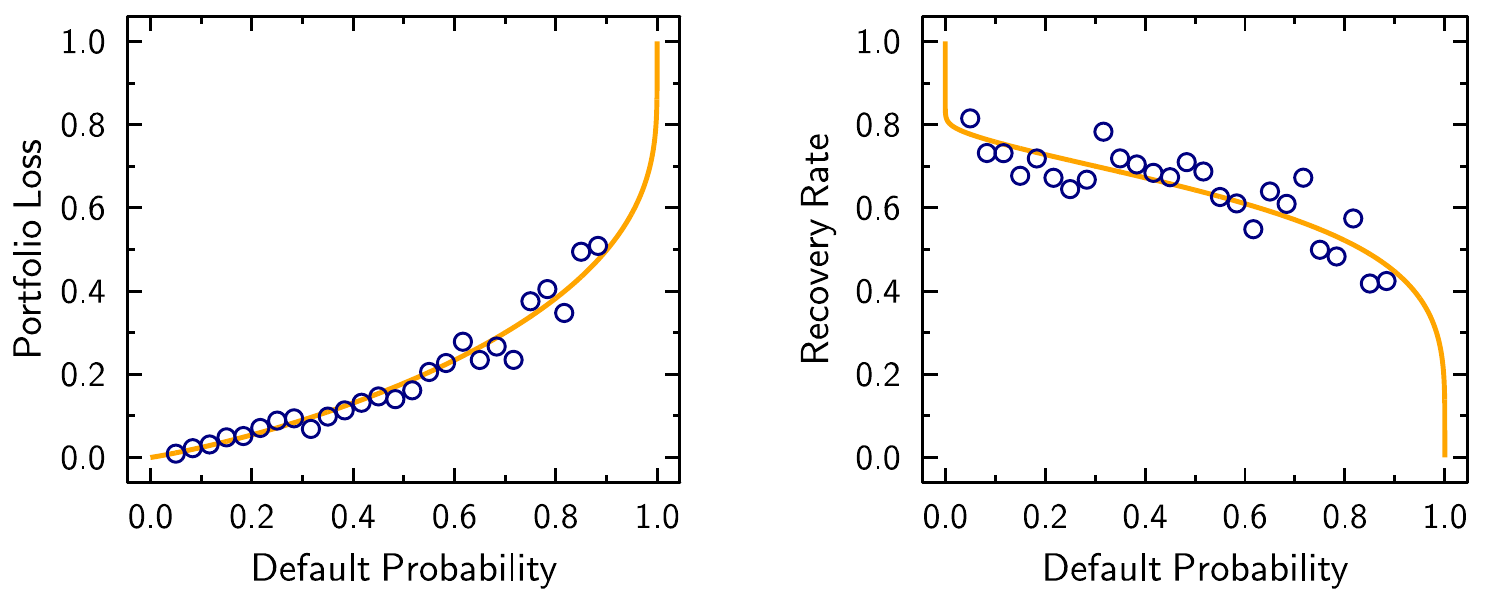}
    \caption{Empirical results from 2000 to 2010 for four-year model-maturities.
    The left plot shows the portfolio loss $\langle \mathrm{L}(\mathrm{PD})
\rangle$, and the right plot shows the recovery rate $\langle
\mathrm{RR}(\mathrm{PD}) \rangle$. The orange line shows the respective
analytical curves with $B = 0.635 $. The parameter has been fitted to the
portfolio loss data.}
    \label{fig:RRLGD4}
  \end{figure}

Note that the recovery rate as calculated in Eq.~\eqref{eq:PDRR} is an average over many realizations of the underlying stochastic process. Accordingly, we can only expect that the data shows the described behavior on average. To account for this, we divide the $\mathrm{PD}$ scale into 30 equally large bins. We then compute the mean of all recoveries in each bin. We disregard bins with less than five data points. The results are presented in Figs. \ref{fig:RRLGD2} and \ref{fig:RRLGD4} 
for two-year and four-year model-maturities, respectively.

We find that the average recovery rate decreases when the probability of default increases. Moreover, the results are well described by Eq.~\eqref{eq:PDRR}.
The parameter $B$ was determined by a least-squares fit of the data in the plot for the portfolio loss $L$. We find $B = 0.882$ and $B = 0.635$ for model-maturities of two years and four years, respectively. We find $B$ to decrease for longer model-maturities, which is not expected from the structural model; Equation \eqref{eq:B} implies $B\sim\sqrt{T}$.

It is worth pointing out that the negative correlation can be found whether the evaluated bonds are senior secured or senior unsecured. 
Also in the latter case, the dependence between default and recovery rates is well described by the structural recovery model.

\section{Conclusions}
\noindent
We have investigated the relation of default probabilities and recovery rates empirically, using Moody's Default and Recovery Database. 
These two measures of credit risk show a strong negative correlation in accordance with literature. 
The structural recovery rate, derived for the Merton model in a previous study, is able to describe the empirical dependence of default and recovery rates.
This is all the more remarkable, as the Merton model makes many unrealistic assumptions; for instance, it neglects coupon payments, infers a very simple capital structure of the debtors, and does not allow for defaults prior to maturity.
However, the strong empirical support for the structural recovery rate encourages to incorporate it into current credit risk models.

\section*{Acknowledgments}
\noindent
We thank Thilo Schmitt and Thomas Guhr for helpful discussions.


\end{document}